\documentclass[12pt]{article}

 \usepackage{amsfonts}
 \usepackage{amssymb,amsmath}
 \usepackage{graphicx} \usepackage{epstopdf}
 \newcommand{\crlb}[1]{\label{#1}\\[2pt]}
 
 \newcommand{\crld}[1]{\label{#1}}
 \newcommand{\eela}[1]{\quad\hbox{\scriptsize{#1}}\label{#1}\end{eqnarray}}
 \newcommand{\eelb}[1]{\label{#1}\end{eqnarray}}
 
 \newcommand{\newsecb}[2]{\section{#1}\label{#2}\setcounter{equation}{0}}

 \newcommand{\nolabels} {\def\eel{\eelb}\def\eeql{\eeqlb}  \def\crl{\crlb} 
 \def\newsecl{\newsecb}\def\bibiteml{\bibitem} \def\citel{\cite}\def\labell{\crld}}
\newcommand{\eeqla}[1]{\quad\hbox{\scriptsize{#1}}\label{#1}\end{aligned}\end{equation}}
\newcommand{\eeqlb}[1]{\label{#1}\end{aligned}\end{equation}}

\newcommand\publishversion  {\nolabels\setlength{\textheight}{8.3in}\setlength
    {\oddsidemargin}{0in} \setlength{\textwidth}{6.3in}\setlength{\topmargin}{-0.2in}}

\def\beq{\begin{equation}\begin{aligned}}		\def\eeq{\end{aligned}\end{equation}}
\def\be{\begin{eqnarray}}  					\def\ee{\end{eqnarray}}		
   \def\bi#1{\begin{itemize}\item[#1]} 	      	   \def\ei{\end{itemize}}

		             \def\t{\tau}    

 	 		\def\s{\sigma}     	      	 
	     		          		              	  
\def\W{\Omega}

          			\def\ra{\rightarrow}

\def\ffract#1#2{\raise .2 em\hbox{$\scriptstyle#1\,$}\kern-.3em/\kern-.2em\lower .15 em \hbox{$\scriptstyle\,#2$}}

\def\bpmatrix{\begin{pmatrix}} 			\def\epmatrix{\end{pmatrix}}
\def\bmatrix{\begin{matrix}} 			\def\ematrix{\end{matrix}} 
\def\bcenter{\begin{center}}			\def\ecenter{\end{center}}

\def\lowerheightfig#1#2#3{\(\raise-#1\hbox{\includegraphics[height=#2]{#3}}\)}
\def\lowerwidthfig#1#2#3{\(\raise-#1\hbox{\includegraphics[width=#2]{#3}}\)}

\def\th{\({}^{\mathrm{th}}\)}

\def\hok{\(\hbox{ }\)}

\def\weglaten#1{}	

\publishversion 

\begin{document}

\begin{titlepage}
 \title{ \LARGE\bf  An unorthodox view on quantum mechanics
\author{Gerard 't~Hooft}
\date{\normalsize
Faculty of Science,
Department of Physics\\
Institute for Theoretical Physics\\
Princetonplein 5,
3584 CC Utrecht \\
\underline{The Netherlands} \\[10pt]
e-mail:  g.thooft@uu.nl \\ internet: 
http://www.staff.science.uu.nl/\~{}hooft101/ }
}
 \maketitle

Quantum mechanics was definitely one of the most significant and important discoveries of 20\th century science. It all began, I think, in the year 1900 when Max Planck published his paper entitled ``On the Theory of the Energy Distribution Law of the Normal Spectrum"\,\cite{Planck1900}. In it, he describes a simple observation: if one attaches an entropy to the radiation field as if its total energy came in packages -- now called quanta -- then the intensity of the radiation associated to a certain temperature agrees quite well with the observations. Planck had described his hypothesis as `an act of desperation'. But it was the only one that worked. Head on. Until that time, the best attempt at performing such a calculation had resulted in the Rayleigh-Jeans law, an important result at the time, but this law only worked at the lowest frequencies of the radiation, whereas it failed bitterly at high frequencies / smaller wavelengths. The Rayleigh-Jeans law would yield a badly divergent, hence meaningless, expression for the intensity of the radiation emitted at high frequencies.

The history of what happened next, is well-known, and has been recounted in excellent reports\,\cite{Pais1986}, which, I think, do not need to be repeated here. 
It was realised that all oscillatory motion apparently comes in energy packets, which seem to behave as particles, and that the converse should also be true: all particles with definite energies must be associated to waves. This all culminated in the \emph{theory of quantum mechanics.} The year 1926 provided a new landmark: Erwin Schr\"odinger's equation\,\cite{Schroedinger1926}. The title of my essay was not intended to cast doubt on this equation; the Schr\"odinger equation has been and still is the pivotal equation on which much of physics and all of chemistry is based, and I am in awe of it just as many researchers before and after me are. 

\end{titlepage}

\setcounter{page}{2}

Indeed, in his original paper, Schr\"odinger went quite far in discussing Hamilton's principle, boundary conditions, the hydrogen atom and the electromagnetic transitions from one energy level to the next. The `confusions' I am referring to arose when more and more researchers began to dispute the question how the equation is supposed to be interpreted. Why is it that positions and velocities of particles at one given moment cannot be calculated, or even defined unambiguously?

We know very well how to use the equation. The properties of atoms, molecules, elementary particles and the forces between all of these can be derived with perplexing accuracy using it. The way the equation is used is nothing to complain about, but what exactly does it say? 

The first question one may rightfully ask, and that has been asked by many researchers and their students, is:\bi{}
\emph{What do these wave functions represent?} In particular the ones that are \emph{not} associated to photons (the energy packets of the electromagnetic field, which we think we understand very well). What do those waves stand for that are associated to electrons, or other elementary particles, or even molecules and larger things, including cats, and eventually, physicists? What happens to its wave function when you actually observe a particle? \ei
One extremely useful observation was made by Max Born\,\cite{Born1926}: The absolute square of a wave function, at some spot in position space, must simply stand for the \emph{probability} to find the particle there. This made a lot of sense, and it was rightly adopted as a useful recipe for dealing with this equation. But then, 
many more questions were asked, many of them very well posed, but the answers sound too ridiculous to be true, and, as I shall try to elucidate, they \emph{are} too ridiculous to be true. The truth is much simpler to grasp conceptually, even if the mathematics needed to back this up can be delicate. This is astonishing. Almost a full century has passed since the equation was written down, and fierce discussions have been held, quite a few standpoints were vigorously defended and equally vigorously attacked. We still do not know what or whom to believe, but it still goes on, while others get irritated by all this display of impotence\,\cite{NGvK1988}. Why is it that we still do not agree?
I think I know the answers, but almost \emph{everyone} disagrees with me.

In fact, one possible reply is that one could decide to ignore the question. Dirac for instance, advised not to ask questions that cannot be answered by any experiment. Such questions are of secondary importance. We know precisely how to use Schr\"odinger's equation; all that scientists have to do is insert the masses and coupling parameters of all known particles into the equation, and calculate. What else can you ask for? Many of my colleagues decided to be strictly `agnostic' about the interpretation, which is as comfortable a position to take as what is was for 19\th century scientists to stay `agnostic' about the existence of atoms.

But yes indeed, there is something else. What \emph{are} those masses and coupling strengths? Do particles exist that we have not yet been able to detect? Isn't it the scientist's job to make predictions about things we have not yet been able to unravel? These are questions that are  haunting us physicists. We have arrived at a splendid theory that accounts for almost anything that could be observed experimentally. It is called the \emph{Standard Model} of the subatomic particles. But this model also tells us that particles and forces may exist that we could not have detected today. Can we produce any theory that suggests what one might be able to find, in some distant future? And as of all those particles and forces that we do know about, is there a theory that \emph{explains} all their details? 

Today's theories give us little to hang on to. This is why it is so important to extend our abilities to do experiments as far as we can, Recently, audacious plans have been unfolded by the European particle physics laboratory CERN, for building a successor for its highly successful Large Hadron Collider (LHC). While the LHC has provided for strong evidence supporting the validity of the Standard Model up to the TeV domain, theoreticians find it more and more difficult to understand why this model can be all there is to describe what happens further beyond that scale. There must be more, but our theoretical reasoning leads to more questions doubting the extent to which this model can be regarded as `natural' when more of the same  particles with higher  energies are included, while the existence of totally new particles would be denied. 

Inspired by what historians of science are telling us about similar situations in the past history of our field, investigators are hoping for a `paradigm shift'. However, while it is easy to postulate that we `are doing something wrong', most suggestions for improvement are futile; suggesting that the Standard Model would be `wrong' is clearly not going to help us. The `Future Circular Collider' is a much better idea; it will be an accelerator with circumference of about 100 km, being able to reach a c.m.~collision energy of 100 TeV. The importance of such a device is that it will provide a crucial background forcing theoreticians to keep their feet on the ground: if you have a theory, it better agree with the newest experimental observations.

Ideally, we should have both experiments \emph{and} theories. 
The Standard Model itself contains more than 20 fundamental parameters, `constants of nature', that are begging for an explanation. The masses of the fermions (`matter particles') are ascribed to the Brout-Englert-Higgs mechanism, just as the mass of the Higgs particle itself. Yet the Brout-Englert-Higgs theory does not tell us why the postulated coupling parameters are such that they generate the mass spectrum that has been measured, numbers that vary wildly. Other interaction parameters (those of the gauge fields, due to the `force carrying particles'), can be explained to some extent, though most details of their origin are also mysterious. What we see is that the relative strengths of the force couplings hint towards a further unification at extremely high energies, though we have not seen any verification of predictions arising from this idea. 
There is definitely something that we have not understood.

A notorious and brave approach is called \emph{superstring theory}. Superstring theory has not lead to new predictions -- apart from ones that were not verified by observations at all, such as the prediction of supersymmetry at LHC energies. The  superstring theory approach has been criticised a lot for this `failure', but it must be emphasised that other approaches achieved far less. We can note that, although the Standard Model was not explained at all, string theory does lead to a zoo of particles that are not so unlike the Standard Model ones, and many of its practitioners interpret this fact as an encouragement to reinforce their efforts. All I can add is that the more we keep staring at the Standard Model's parameters, the more it seems that natural explanations should be asked for, and the existing theories do not seem to converge towards answers.

It is here that I suggest to look at the quantum mechanical nature of all observed phenomena.  Superstring theory has not provided for any explanation as to what quantum mechanics is about.
None of today's cherished theories are indicating any evidence for conceivable constraints on the quantum mechanical nature of their equations.

Imagine that further scrutinising what is called quantum mechanics could lead to new ideas. Perhaps there are things that we don't see yet but that we can speculate about. 
As for quantum mechanics, let me briefly remind the reader what the developments were that got us entwined into a messy knot of problems.  Back in 1932, John von Neumann gave a ``proof"\,\cite{Neumann1932} of the statement that quantum mechanics cannot be explained in terms of 
 ``hidden" degrees of freedom, which would restore some deterministic interpretation of Schr\"odinger's wave function. His proof was later dismissed. Both the proof and its dismissal showed that the authors understood the mathematical nature of the equation, but missed the point that the physical world might be much more complex than just this one abstract equation, in which all sorts of possible complications were suppressed.

A very important paper was written by Einstein, Podolsky and Rosen\,\cite{EPR1935}, arguing that indeed some essential ingredient was missing in this quantum theory. In a nut shell:  if you consider an atom emitting two photons at once, quantum mechanics suggests that you can measure the position of one photon and the momentum of the other photon, but since the two photons are equal (`entangled', to be precise), these measurements together should give you information for both photons that would be forbidden by the same quantum theory. 

In 1952, D.~Bohm took an original idea of L.~de Broglie literally: ``the wave function acts as a pilot, telling the particle where to go"\,\cite{Bohm1952}. 

John S. Bell took up the same question\,\cite{Bell1964}, but insisted on restoring the notion of locality and causality in the interpretation of the wave function; both Bohm's theory and the EPR argument describe the events while giving up the notion of locality. Bell tried to rephrase everything in a powerful theorem, which would become famous. Starting from assumptions that he considered to be utterly reasonable, he derived: \emph{you can't build a theory of hidden variables that explains what Schr\"odinger's equation describes, if you insist that it be local.}

A different avenue was the argument that the wave function just displays \emph{everything} that might be happening, including all alternatives: an infinite set of different universes can all be evolving, together forming an even grander body later called the `multiverse', or `omniverse'. This multiverse would be like a gigantic tree, every branch, every leaf of it being an entire universe. This `many world theory' was formulated by Hugh Everett\,\cite{Everett1957} and strongly advocated by Bryce S.~DeWitt\,\cite{DeWitt1973}. This theory would not change the equations, and basically, it would take for granted that quantum mechanics can never predict what will happen, beyond the level of statistical statements. You will always be surprised as to which world you apparently entered into, while, with different probabilities, all other possible worlds are realised somewhere else in the multiverse.

I think they are all mistaken. What the Schr\"odinger equation is describing is not exactly what is happening; it merely describes the tip of a gigantic iceberg, in which most processes happen far beneath the water line. All that needs to be realised is that, like in \emph{any} scientific theory, our predictions concerning the outcomes of experiments cannot be expected to be infinitely accurate. There are margins of error, due to the fact that we have been using statistics. The statistical assumptions that went \emph{into} Schr\"odinger's equations are about variables that give shape to space and time, fluctuating far too fast to be included in such a way that we can control them. We don't control them and this is the reason why we discover that our predictions come with margins of error, like in any other scientific theory. I think that the internal mathematical nature of Schr\"odinger's equation supports this view, but there is too much that we do not know at present, so that building models for what ``really" happens is still difficult for us.

This idea indeed would be covered by the name ``hidden variables", but the hidden variables are far more sophisticated than what the earlier researchers had been thinking of -- I return to the consequences  of our use of statistics shortly.

Some of my readers will object that Bell did emphatically include such hidden variables in his treatment, but Bell made assumptions that need not be valid. Indeed, what Von Neumann had done, and what Bell did later, was to construct a so-called `no-go theorem'. They both badly wanted to construct a hidden variable theory, but when they did not succeed, they decided to close the lid for good, that is, to \emph{prove} that such a theory cannot be constructed. Time and again, however, history of science has shown that no-go theorems do not say much more than that the avenues inspected by their authors do not lead to a desired result, so that, all that should have been done is modify the assumptions.

To obtain a more solid understanding of what quantum mechanics is really telling us, we have to go back a long way into the history of physics as a science. What happened was that long sequences of discoveries not only enriched our knowledge tremendously, they also narrowed down our way of thinking, so much so that some obvious facts are now considered to be obsolete, just because, in practice, our streamlined vision has become much more powerful. Take for instance the way we describe a moving object. We use \emph{real numbers} to describe positions, velocities, energies and so on. What are real numbers? They are a natural generalisation of `rational numbers'. Rational numbers are what you get if you divide an integer by an other integer. \emph{Integers} are numbers that you get simply by counting, for instance the number of steps needed to get across your room. Integers can be stored in a computer, but before doing so we have to realise that there are limits to how far we can count, in practice. The simplest integers are called `bytes'. They can vary from 0 to 255 but no further than that. Each byte can be expressed in terms of eight `bits'. Bits can take only two values, which you can call 0 and 1, or alternatively, \hbox{+ and --}.

Those real numbers used now, are so powerful that we forget that they are man-made. Every real number that we use to describe our world would require an infinite amount of computer space to store it, because it contains an infinite sequence of digits, all of which can be arbitrary. Is this really the correct way to indicate distances, velocities and so on? It could be that all positions and all velocities eventually only require integers to specify their values. This would mean that particles live in a space-time that is not strictly continuous but could be what is called a \emph{lattice}.

I am not saying that I `know' that space-time is a lattice. I don't, but I do know that possibilities such as this would profoundly change the ways one could think of a `particle', and with that, the way we should formulate the laws of physics they obey. In practice it is very hard to formulate laws of physics for lattice-particles that fit well with the tremendously sophisticated science that we have today. Schr\"odinger's beautiful equation seems to be what you get if you want to describe particles but at the same time you want to hang on to those, somewhat artificial, real numbers.

There is another aspect of modern science that has become so self-evident that researchers fail to notice that the deeper physical laws could deviate from standard dogma. It is called `statistics'. Quantum mechanics gives us ways to estimate very precisely the statistical behaviour of large numbers of particles, or large numbers of experiments, or large numbers of observers. When you consider a large reservoir containing molecules or other particles moving chaotically, then it is only too tempting to say that the whole lot of them will be fundamentally unpredictable when we ask for all details of the motion of every one of the particles individually. There will be exceptional chance events, and they will be fundamentally unpredictable, with or without quantum mechanics.

To formulate his theorem, John Bell needed to introduce two hypothetical observers called Alice and Bob. Alice and Bob both had to choose what they would measure. They had  the \emph{free will} to decide what to choose. But if we would have an omnipotent theory that prescribes every move with infinite precision, such a free will does not exist. Nobody today thinks of physical laws that also control the outcomes of arbitrary, individual, measurements in a gas consisting of billions of particles, but that \emph{is} what we should try to find. Note that I am not pretending that a theory can be found that `predicts' where every single atom or molecule will be going, but I do think that it is reasonable to suspect that laws of nature exist that would uniquely fix their behaviour, while our ability to predict will always be limited by our knowledge, our computation power, and our understanding.

To many of my readers (the ones who may still be with me), what I just said sounds very much like letters we receive in our daily mail from amateur physicists. They are amateurs because they usually exhibit a dismal lack of knowledge and understanding of modern science. Like many of my colleagues, I quickly discard such letters, but sometimes they are fun to read. More to the point, by not knowing how our world has been found to hang together, they could have bounced into some more independent ways of asking questions. Indeed, when I think of questions concerning quantum mechanics, I know I have to make giant leaps of logic. Often, these are giant leaps backwards. What I really want is use kinds of logic that do not make use of conventional number systems, or other conventional ways to describe our understanding of physical laws. But then, before irritating my colleagues with my `crackpot' findings, I search for the proper connections with real science. This is because I do know what conventional science says about quantum mechanics, particles, and number systems. And today I think of ways to attach unconventional views to the beautiful edifice called {the Standard Model}.

This is the reason why the discussion about quantum mechanics may be an important discussion after all. We can't accept a theory that, in the very end, only predicts statistical properties of the small particles, but leaves them free, to some extent, to do whatever they like. If you dig deep enough, you should find some superior scenario of the dynamical laws, and this new scenario must carry implications for the way we build models.

The Standard Model does not pose any restrictions concerning the quantum equations on which it is based, but exactly that might be what we need, to make the next step in particle theory. Today, the known elementary particles form a bunch of objects with masses ranging from a few milli-electronVolts (meV) to somewhere around a TeV (tera-electronVolts), a stretch of some 15 orders of magnitude.  Particles heavier than a few TeV cannot be observed with the machines we have today. There is no fundamental limit to the mass of a particle until we reach the Planck mass, some 16 orders of magnitude beyond where we are now. It is the meshes of our lattice that forbid elementary objects beyond that. This is my iceberg: only one part in \(10^{16}\) is visible above the waterline. 

The particles hidden from our view today, would undergo interactions and oscillations that are so much faster as time proceeds, that we may call them `fast hidden variables'. Today's Schr\"odinger equation has all fast variables suppressed; we cannot detect such fast phenomena. Now, I claim that the fast, invisible phenomena could be described by classical laws, but these movements are fundamentally unobservable, and we can only deal with what we see. In a way invisible to us now, the things we see are controlled by fast moving particles, particles that form the bulk of the iceberg. Today's science gives us an equation that can be used to evade those particles, but this has come at a price: we have hit upon unpredictable behaviour that we should not have been surprised about: these are the whims of the hidden variables. To my taste, this way to interpret what the quantum mechanical laws are about, sounds much more reasonable than assumptions concerning `pilot waves' or `multiverses'. But most importantly, \emph{any} reference to statistical averaging should be banned from the initial assumptions because these indeed lead to no-go theorems such as Bell's. Experimenters do not have the `free will' to do something that is not controlled by deterministic laws of nature.

When making such claims, one has to come with mathematical models. The models we have today are imperfect, but encouraging indications exist that the Schr\"odinger equation is exactly what one should expect from such hidden physical processes\,
\cite{GtH2020, GtH2021}.

I actually found that it may be important to make yet another step backwards: we have become used to the utility of \emph{complex} numbers to describe quantum wave functions. Why? Some people claim it to be one of those quantum peculiarities, setting quantum mechanics aside from all classical theories, even though complex numbers have often also proved useful for calculations in classical topics. What is a complex number? One of the marvels of modern science is that we learned how to take the square root of negative real numbers, but then here also, understanding is often replaced by mysticism. 

This is when one forgets how complex numbers are defined. Complex numbers are simply pairs of real numbers, with rules attached on how to add, subtract, multiply and divide such pairs of numbers. In quantum theories, one is tempted to forget what this means: the physical state of a system includes one or more objects that can come in two forms, represented by the two components of the complex function that represents a wave function. One can also decide to describe such components in a more direct manner, avoiding the use of complex quantities, which after all, are also man-made inventions. Such a direct language may lead to some useful insights. When restricting ourselves to real wave functions only, one finds that the quantum Hamiltonian is an antisymmetric (real) matrix. Finding eigenvectors and eigenvalues of matrices is common practice for easing our calculations, but here, antisymmetry would force us to split the states into pairs again, leading us straight back to those useful complex numbers. 

The mathematical insights gained through this brief excursion may turn out to be very useful.     Quantum mechanical models with a deeper, deterministic basis  cannot support the use of arbitrary real numbers as input parameters. This means that constants such as masses and coupling strengths of particles cannot be assumed to be any real number. In some intricate way, these must be expressed in terms of integer numbers  only. This suggests that one might hit upon special types of field theories where the input parameters can be calculated.

 \end{document}